\documentclass[lathuile]{article}
\usepackage{amssymb}
\usepackage{amsmath}
\usepackage{graphicx}
\usepackage{amscd,multicol}

\title{\bf EXTRA GENERATIONS, DISCREPANCIES OF ELECTROWEAK PRECISION DATA AND MASS OF THE HIGGS}
\author{V.A. Novikov\\
{\it ITEP, Moscow, Russia}  }
\date{}
\begin{document}
\maketitle

\begin{center}

Abstract

\end{center}

The latest electroweak precision data are analyzed assuming the
existence of the fourth generation of leptons ($N, E$) and quarks
($U, D$), which are not mixed with the known three generations. If
all four new particles are heavier than $Z$ boson, quality of the
fit for the one new generation is as good as for the Standard
Model. In the case of neutral leptons with masses around 50 GeV
(``partially heavy extra generations'') the  minimum of $\chi^2$
is between one and two extra generations. The predicted value of
the higgs mass $m_H$ in the presence of the fourth generation can
be made rather large. The quality of fits drastically improves
when the data on b- and c-quark asymmetries and new NuTeV data on
deep inelastic scattering are ignored.

\newpage

My talk rests on the results of two papers written recently in collaboration with
L.Okun, A.Rozanov and M.Vysotsky \cite{3a,3b}.

{\bf A Brief History of the SM fits }

For more than ten years the Standard Model (SM) enjoyed  solid agreement
with precision  data provided by experiments at LEP, SLC and  Tevatron.

The matter came to a climax at the time of La Thuille 2000 conference when the SM
fit of the whole set of available electroweak precision data
became absolutely perfect: $\chi^2/n_{\rm d.o.f.} \approx 15/14$.
From the  point of view of Statistics no one sample
of New Physics could improve such fit - it could make it worse or of the
same quality in the best case.

For example, in paper \cite{1} written at that time we reanalyzed the
non-decoupled New Physics in a form of additional heavy
quark-lepton generations.\footnote{ In this extension of the SM leptons of fourth
generation (E,N) should be very weakly mixed with the ordinary
ones, while in quark sector (U,D) mixing is limited only by
unitarity of $3\times 3$ CKM matrix. Implications of
extra quark-lepton generations for precision data were studied in
a number of papers \cite{4a} - \cite{He}.} We confirmed that in the
case of all four new fermions ($U$ and $D$ quarks, neutral lepton
$N$ and charged lepton $E$) heavier than $Z$ boson the radiative corrections to
low-energy observables
were large and the quality of the fit dropped down. As a result such extension of the SM was
 excluded by the data. In particular we found that one heavy generation was excluded at
2.5 $\sigma$ level. We also found that corrections due to existence of relatively
light neutral lepton $N$ ($m_N \approx 50$ GeV) and corrections due to heavy
$U$, $D$ and $E$ could compensate each other and that the SM with additional "partially heavy"
generation is allowed by presicion measurements of low-energy observables as well as the SM itself.
This was an example of the conspiracy of New Physics.

From that time situation with the quality of the SM fit has been changed.
At the time of Osaka Conference (summer 2000) some of the central
values of observables
have been shifted within one sigma, some of the error bars have become slightly smaller.
Nothing radical happened with any of observables but the coherent result was
quite visible and the SM fit
became less good: $\chi^2/n_{\rm d.o.f.} = 21/13$. The level
at which one extra heavy generation was excluded went down to
$2\sigma$ \cite{2}.

For the latest precision data (summer 2001)  \cite{3} the SM fit became even worse
$\chi^2 = 24/13$. As for the fit of the SM with one additional heavy generation
it became approximately of the same quality as for the SM.

To see that we  present in Fig.1 the exclusion plot for the number $N_g$ of extra heavy generations.

\begin{figure*}[t]
\centering
\includegraphics[width=0.56\textwidth,height=0.34\textheight]
{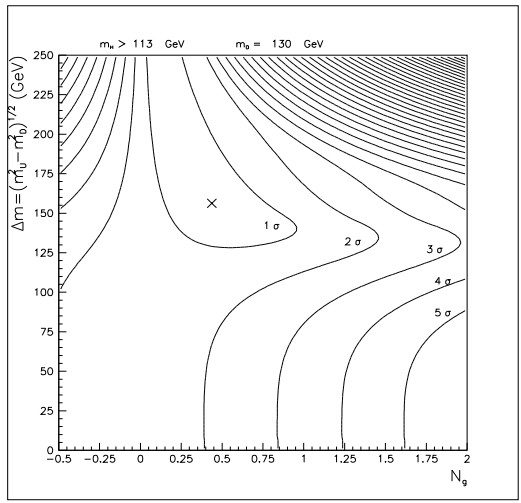} \caption{\it {\label{FIG1} Exclusion plot
for heavy extra generations with the input: $m_D = m_E = 130$ GeV,
$m_U = m_N$. $\chi^2$ minimum shown by cross corresponds to
$\chi^2/n_{d.o.f.} = 22.2/12$, $N_g = 0.4$, $\Delta m = 160$ GeV,
$m_H = 116$ GeV. $N_g$ is the number of extra generations. Borders
of regions show domains allowed  at the level $1\sigma, 2\sigma$,
etc. They correspond to $\Delta \chi^2 = 1, 4, 9, 16,$ etc.}}
\end{figure*}

To produce this plot we take $m_D = 130$ GeV -- the Tevatron lower
bound on new quark mass; we use experimental 95\% C.L. bound on higgs
mass $m_H > 113$ GeV \cite{3} and vary $\Delta m = \sqrt{m_U^2 -
m_D^2}$ and number of extra generations $N_g$. (In order to have
two-dimensional plot we arbitrary assumed that $m_N = m_U$ and
$m_E = m_D$; other choices do not change the obtained results
drastically, see discussion below). We see that $\chi^2$ minimum
corresponds to unphysical point $N_g
= 0.5$. For $170$ GeV $< m_U < 200$ GeV we get the same quality of
fit in the case $N_g =1$ as that for the SM ($N_g
=0$). \footnote{In  ref. \cite{55} one can find a statement that extra
heavy generations
are excluded by the recent precision electroweak data. However, analysis
performed  in \cite{55}
 refers to upper and lower parts of Fig. 1, $\Delta m > 200$ GeV
 and $\Delta m =0$, where the
existence of new heavy generations is really strongly suppressed. This
is not the case for the central part of Fig. 1 ($\Delta m \approx
150$ GeV).} We see that both fits are rather bad but
they are equally bad. New Physics  does not make fit worse as compared with the SM.

Two heavy generations are excluded at more than $3\sigma$ level.
Nevertheless, two and even three ``partially heavy'' generations
are allowed when
neutral fermions are relatively light, $m_N \simeq 55$ GeV (see
Fig. 2). \footnote{Using all existing LEP II statistics on the reactions
$e^+ e^- \to \gamma + \nu \bar\nu, \gamma +N \bar N$ in dedicated
search (see \cite{444})one can exclude 3 ``partially heavy'' generations which
contain such a
light $N$ at a level of $3\sigma$ (see \cite{4}), while one or
even two such generations may exist.}
\begin{figure*}[]
\centering
\includegraphics
[width=0.56\textwidth,height=0.34\textheight]
{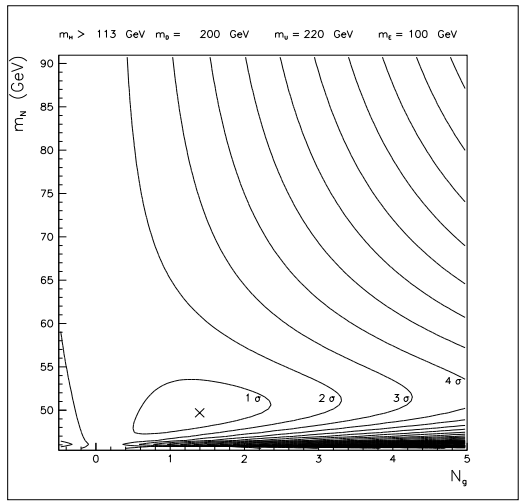} \caption{\it {\label{FIG2} Exclusion plot
for the number of partially heavy extra generations with light
neutral lepton $N$. On horizontal axis the number of extra
generations $N_g$, on vertical axis -- the mass of the neutral
lepton $m_N$. The input: $m_U = 220 $ GeV, $ m_D = 200$ GeV, $m_E
= 100 $ GeV. At the minimum $\chi^2/n_{d.o.f.} = 21.6/12$, $N_g =
1.4$, $m_N = 50$ GeV, $m_H = 116$ GeV. According to LEP
experimental data  $m_N
> 50$ GeV at 95\% c.l. \cite{4, 444}.}
 }
\end{figure*}

Thus the SM fit is bad, the SM with additional heavy (or partially heavy) generations  fit
is also bad. In the literature one can find different modifications of the SM that resolve
one or another discrepancy with experimental numbers  but it seems that not a single modification 
could fit the whole set of data well.
In such situation it is useful to reveal the roots of the bad quality of the SM fit.

{\bf The roots of the SM fit troubles}.

We see three main discrepancies in the existing data.

{\bf Discrepancy 1}.

There is
discrepancy between the average value of $s_l^2$ extracted from the
pure leptonic measurements and its value from events with hadrons
in final state \cite{3}:
\begin{eqnarray}
 & s_l^2 \nonumber \\
 {\rm Leptons} & 0.23113(21) \nonumber \\
 {\rm Hadrons} & 0.23230(29)
\label{1}
\end{eqnarray}
This 3.3 $\sigma$ difference is one of the causes of poor quality of
the SM fit.

The value of hadronic contribution to $s_l^2$ in (\ref{1})
is dominated by very
small uncertainty of  the forward-backward asymmetry
in reaction ~~ $e^+ e^-\to Z\to b\bar b$, measured at LEP
\begin{equation}
(A_{FB}^b)_{\rm exp} = 0.0990(17)
\label{2}
\end{equation}

{\bf Discrepancy 2}.

 There is discrepancy (indirect) between
this LEP result and SLC data. Indeed the value of $A_{FB}^b$
can be calculated by multiplying  beauty asymmetry
$A_b$ and leptonic asymmetry $A_l$ (both measured at SLAC). Then
\begin{equation}
A_{FB}^b = \frac{3}{4} A_b A_l = 0.1038(25) . \label{3}
\end{equation}
The number (\ref{3}) differs from (\ref{2}). Thus
 there is contradiction between LEP and SLC experimental
data.
Moreover SLC number nicely  coincides
with the SM fit: $(A_{FB}^b)_{SM} = 0.1040(8)$ (see e.g. Table 1 from  \cite{3a}).

{\bf Discrepancy 3}.

A new result for
$s_W^2(\nu N)$ and hence for $m_W(\nu N)$ was published by NuTeV
collaboration \cite{11b}:
\begin{equation}
s_W^2(\nu N) = 0.2277(17) \;\; ,
m_W(\nu N) = 80.140(80) \;\; .
\label{4}
\end{equation}
The new value of $m_W(\nu N)$ differs from $m_W$ measured at LEP
II and previously at Tevatron by 3.7 $\sigma$. With new NuTev result we get for the SM fit:
\begin{equation}
m_H = 86^{+51}_{-32} \; {\rm GeV} \;\; ,
\chi^2/n_{d.o.f.} = 30.3/13 .
\label{5}
\end{equation}

 At that moment one can stop and wait for the better data that do not
contradict each other. We are not going to do that, we are going to
proceed with our analisis.

As a guide we take {\bf Lev Landau} advice to young theorists.
According to folklore it sounds like that:

{\bf "Look at the data and multiply experimental errors by factor 3"}

In general case this advice seems too radical.
But in the contradictory situation it seems reasonable to disregard some of the data
to understand their relative contribution into trouble.
The previous consideration demonstrates that the accuracy of $A_{FB}^b $  and new NuTeV data
are under suspicion.
Thus at that point we assume (following Chanowitz \cite{Ch}) that $A_{FB}^b$
has larger uncertainty than given in Eq. (\ref{2}).
If we multiply experimental
uncertainties of $A_{FB}^b$ and $A_{FB}^c$ (which are strongly
correlated) by a factor 10 and do the same with new NuTeV data, the quality of SM fit improves
drastically: $\chi^2/n_{d.o.f.}$ shifts from 23.8/13 to
10.9/13.
(Landau factor 3
leads to a more or less the same result).

However, a new problem arises after removing $A_{FB}^{b,c}$.
It was  known for a long time that the SM fit
results in prediction of light higgs - the central value of its mass was below the
direct lower limit by LEP II. For example in ref.(\cite{3a}) we got for the SM fit that

\begin{equation}
m_H=79^ {+47}_{-29}\; {\rm GeV} \;\; , \label{6}
\end{equation}
It is slightly less than one sigma away from
114.1 GeV bound of LEPII. (The discrepancy is smaller
 in case of inclusion of the
new NuTeV result, see Eq. (\ref {5})). Thus we have one sigma deviation
of the predicted value of higgs mass from the direct LEPII bound. This discrepancy is
not too bad, but the $\chi^2$ of the SM
fit is rather
bad.

With our
modification of experimental results on ${\nu}N$ scattering and on $A_{FB}^{b,c}$
the SM fit gives :
\begin{equation}
m_H = 42^{+30}_{-18} \; {\rm GeV} \;\; , \label{7}
\end{equation}
with good $\chi^2=10.9/13$, but well below modern LEP II bound.
Increasing $m_H$ to the LEPII bound leads to an increased $\chi^2=14.5$,
thus the difference $\delta \chi^2=14.5-10.9=3.6$ is close to 1.9 $\sigma$.
Two sigma difference is not yet a discovery of the violation of SM,
but it is a substantial
trouble for the SM.

Fortunately there are  ways to avoid this trouble.
One possible way to raise the
predicted value of $m_H$ is to assume the existence of fourth
generation of leptons and quarks \cite{3b, He}.

It was noticed
in \cite{He} that the predicted mass of the higgs could be as
high as 500 GeV. That conclusion was based on a sample of 10.000
random inputs of masses of fourth generation leptons and quarks.
In \cite{3b} we used
 our LEPTOP code \cite{33b} to find steep and flat
directions in the five-dimensional parameter space: $m_H$, $m_U$,
$m_D$, $m_E$, $m_N$.
For each point in this space we performed three-parameter fit
($m_t, \alpha_s, \bar{\alpha}$) and calculated the $\chi^2$ of the fit.

It turns out that the $\chi_{\rm min}^2$
depends weakly on $m_U +m_D$ and $m_H$, while its dependence on
$m_U - m_D$, $m_E$ and $m_N$ is strong. Therefore to present the result of
the complete analysis of the Summer 2001 precision data it is enough
to have a few two-dimensional plots. In Figures 3-6 we show
$\chi_{\rm min}^2$ (crosses) and constant $\chi^2$ lines
corresponding to $\Delta \chi^2 = 1, 4, 9, 16,$ ... on the plane
$m_N, m_U -m_D$ for fixed values of $m_U +m_D = 500$ GeV, $m_H =
120$ (Figs. 3 and 5) and 500 GeV (Figs. 4 and 6) and $ m_E=100$
(Figs. 3 and 4) and $300$ GeV (Figs 5 and 6).

\begin{figure*}[]
\centering
\includegraphics[width=0.64\textwidth,height=0.4\textheight]{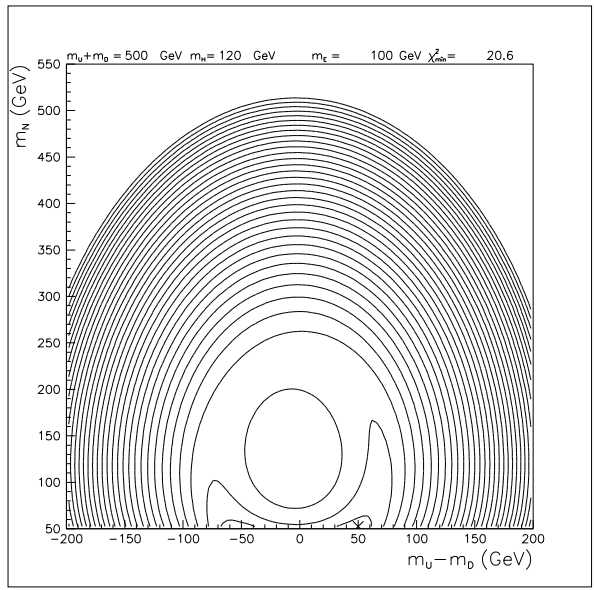}
\caption{\it {\label{FIG3} Exclusion plot on the plane $m_N, m_U
-m_D$ for fixed values of $m_H=120$ GeV,
 and $m_E=100$ GeV.
 $\chi_{\rm min}^2$ shown by two crosses corresponds to $\chi^2/n_{d.o.f.} =
20.6/12$. (The left-hand cross is slightly below $m_N = 50$ GeV.)
 The plot was based on the old NuTeV data. The new NuTeV data preserve
the pattern of the plot, but lead to $\chi_{\rm min}^2/n_{d.o.f.}
= 27.7/12$. If  $A_{FB}^b$ and $A_{FB}^c$ uncertainties are
multiplied by factor $10$ we get $\chi_{\rm min}^2/n_{d.o.f.} =
19.1/12$ for  new NuTeV, and $\chi_{\rm min}^2/n_{d.o.f.} =
11.3/12$ for old NuTeV.}}
\end{figure*}

\begin{figure*}[]
\centering
\includegraphics[width=0.64\textwidth,height=0.4\textheight]{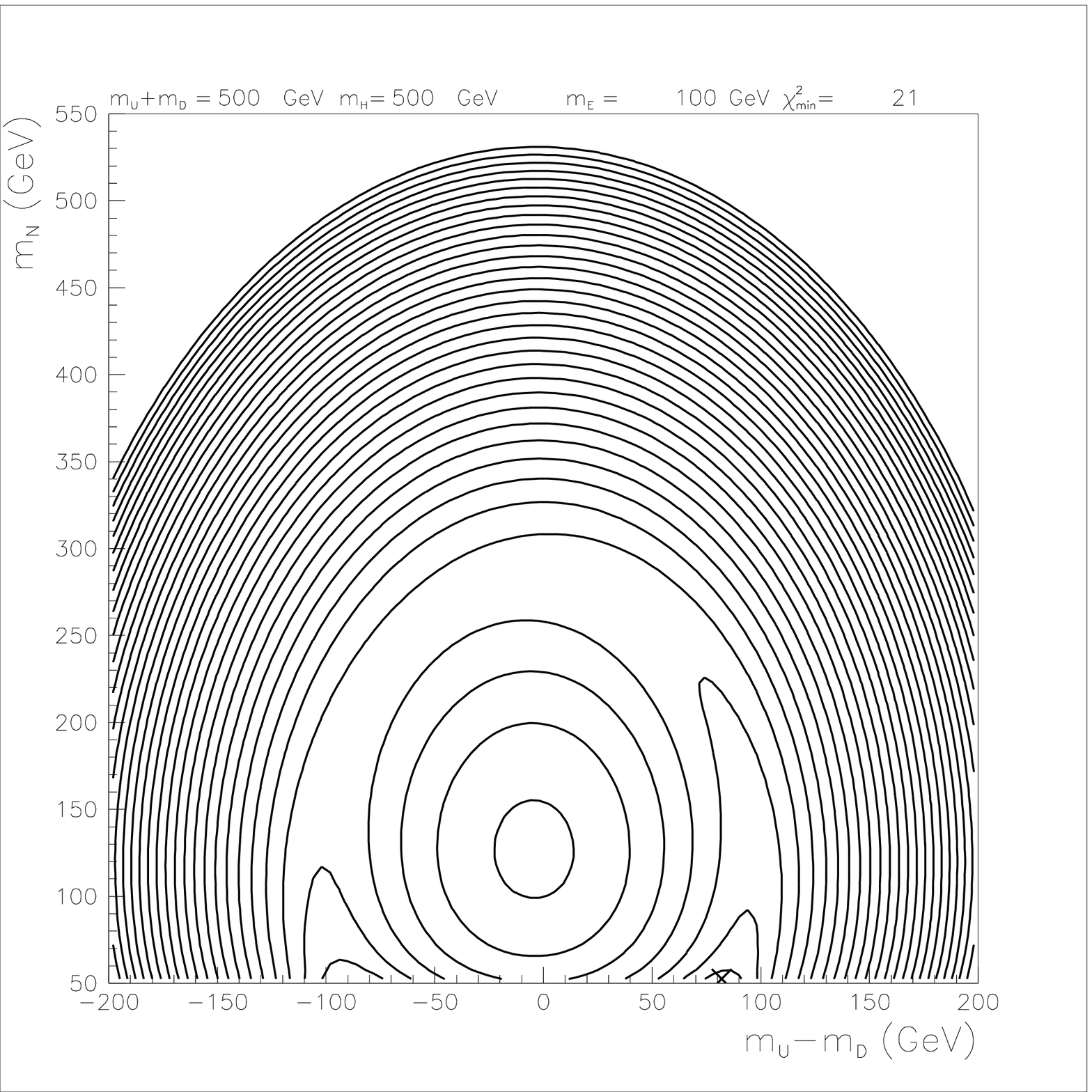}
\caption{\it {\label{FIG4}  Exclusion plot on the plane $m_N, m_U
-m_D$ for fixed values of $m_H=500$ GeV, and $m_E=100$ GeV.
 $\chi_{\rm min}^2$ shown by two crosses corresponds to $\chi^2/n_{d.o.f.} =
21.4/12$. (The left-hand cross is slightly below $m_N = 50$ GeV.)
The plot was based on the old NuTeV data. The new NuTeV data
preserve the pattern of the plot, but lead to $\chi^2_{\rm
min}/n_{d.o.f.} = 28.3/12$. If  $A_{FB}^b$ and $A_{FB}^c$
uncertainties are multiplied by a factor $10$, we get $\chi^2_{\rm
min}/n_{d.o.f.} = 21.2/12$ for  new NuTeV, and $\chi^2_{\rm
min}/n_{d.o.f.} = 13/12$ for old NuTeV.}}
\end{figure*}

\begin{figure*}[]
\centering
\includegraphics[width=0.64\textwidth,height=0.4\textheight] {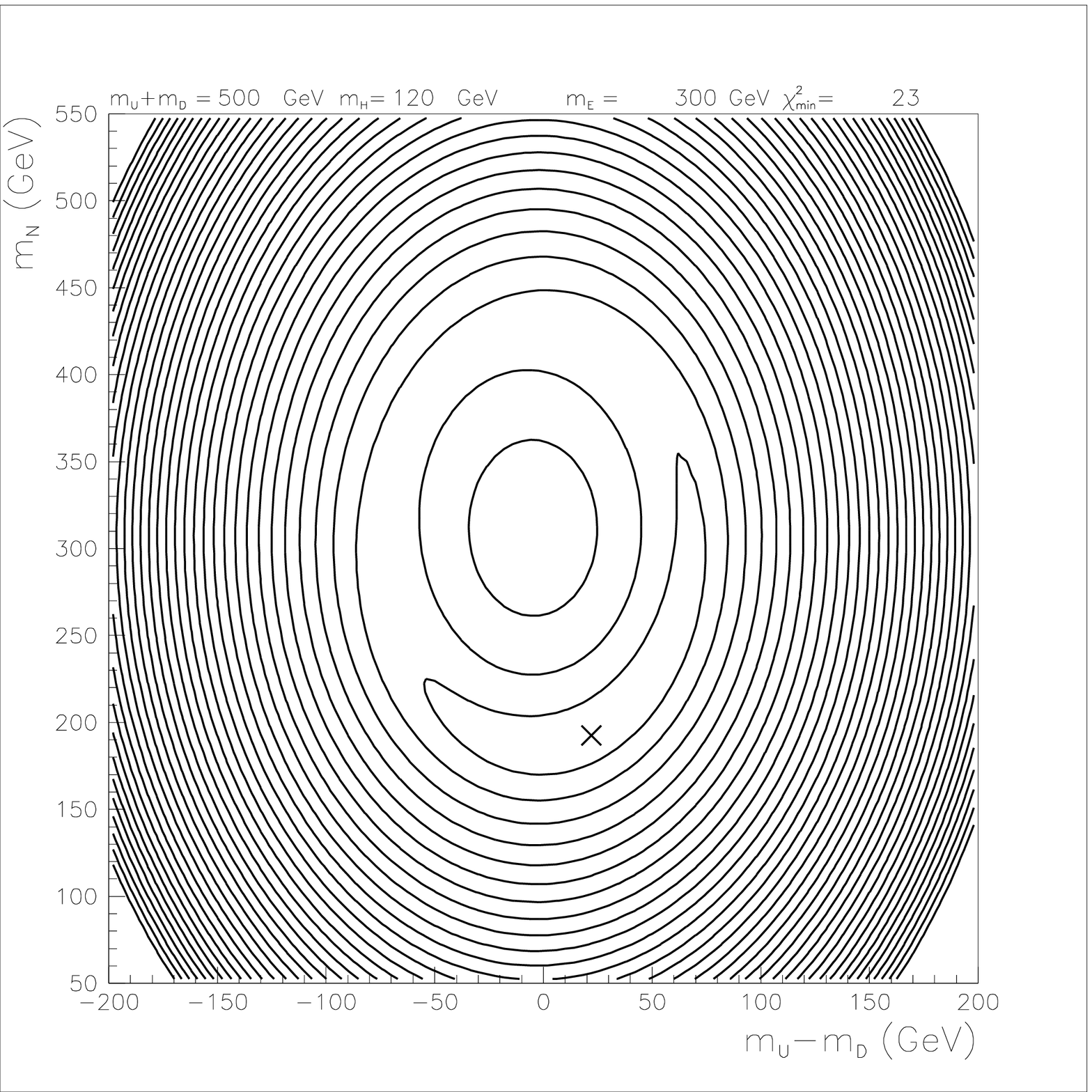}
\caption{\it {\label{FIG5} Exclusion plot on the plane $m_N, m_U
-m_D$ for fixed values of $m_H=120$ GeV, and $m_E=300$ GeV.
 $\chi_{\rm min}^2$ shown by cross corresponds to $\chi^2/n_{d.o.f.} =
23.0/12$.
 }}
\end{figure*}

\begin{figure*}[]
\centering
\includegraphics[width=0.64\textwidth,height=0.4\textheight]{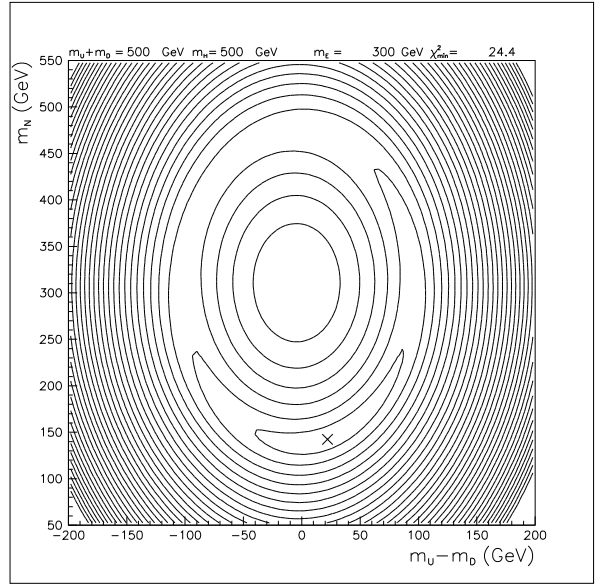}
\caption{\it {\label{FIG6} Exclusion plot on the plane $m_N, m_U
-m_D$ for fixed values of $m_H=500$ GeV, and $m_E=300$ GeV.
 $\chi_{\rm min}^2$ shown by cross corresponds to $\chi^2/n_{d.o.f.} =
24.4/12$.
 }}
\end{figure*}

The above choice of masses is based on a large number of fits
covering a broad space of parameters: 300 GeV $< m_U + m_D <$ 800
GeV; 0 GeV $< m_U - m_D <$ 400 GeV; 100 GeV $< m_E <$ 500 GeV; 50
GeV $< m_N <$ 500 GeV;  120 GeV $< m_H <$ 500 GeV. Concerning
quarks, $m_U + m_D$ is bounded from below by direct searches
limit, while from above by triviality arguments. Since $\chi^2$
dependence on $m_U + m_D$ is very weak, our choice of intermediate
value $m_U + m_D = 500$ GeV represents a typical, almost general
case. For this choice $|m_U - m_D|$ can not be larger than $\sim
200$ GeV because of the mentioned above direct searches bound.

Concerning charged lepton, its mass is taken above LEP II bound.
We present fits at two values of $m_E$ (100 GeV and 300 GeV) and
one can see how fit is worsening with $m_E$ going up.

Concerning the value of $m_H$, we vary it from the lower LEP II
limit up to triviality bound and since the dependence of
observables on $m_H$ is flat, one can get $\chi^2$ behavior from
two limiting points: $m_H = $120 and 500 GeV.

For $m_E=100$ GeV we have the minimum of $\chi^2$ at
 $m_N \simeq 50$ GeV and:

\begin{center}
\begin{tabular}{lll}
for $m_H = 120$ GeV: & $|m_U -m_D| \sim 50$ GeV, &
$\chi_{\rm min}^2/n_{d.o.f.} = 20.6/12$  \\
for $m_H = 300$ GeV: & $|m_U -m_D| \sim 75$ GeV, &
$\chi_{\rm min}^2/n_{d.o.f.} = 20.8/12$ \\
for $m_H = 500$ GeV: & $|m_U -m_D| \sim 85$ GeV, &
$\chi_{\rm min}^2/n_{d.o.f.} = 21.4/12$
\end{tabular}
\end{center}

Thus we have two lines ($m_U>m_D$ and $m_U<m_D$) in the ($m_U -
m_D, m_H$) space that correspond to the best fit of data. Along
these lines the quality of the fit is only slightly better for
light higgs ($m_H \sim 120$ GeV) than for the heavy one ($m_H
\sim$ 300 -- 500 GeV).\footnote{Note that the $n_{d.o.f}$ is 12, unlike the case of the Standard
Model, where it was 13 (ref. \cite{3a}). This change occurs because
in the present paper $m_H$ is not a fitted, but a fixed parameter
(hence 13 becomes 14), while $m_N$ and $m_U - m_D$ are two
additional fitted parameters (hence 14 becomes 12).}

For $m_E=300$ GeV we have the minimum of $\chi^2$ at
 $m_U-m_D \simeq 25$ GeV and:

\begin{center}
\begin{tabular}{lll}
for $m_H = 120$ GeV: & $m_N \sim 200$ GeV, &
$\chi_{\rm min}^2/n_{d.o.f.} = 23.0/12$  \\
for $m_H = 300$ GeV: & $m_N \sim 170$ GeV, &
$\chi_{\rm min}^2/n_{d.o.f.} = 24.0/12$ \\
for $m_H = 500$ GeV: & $m_N \sim 150$ GeV, &
$\chi_{\rm min}^2/n_{d.o.f.} = 24.4/12$
\end{tabular}
\end{center}

Thus, the best fit of the data corresponds to the light $m_E
\simeq 100$ GeV and $m_N\simeq 50$ GeV. The significance of light
$m_N$ (around 50 GeV) was first stressed in \cite{1}. Increase
of $m_E$ leads to the increase of $m_N$ and to fast worsening of
$\chi_{\rm min}^2$.

Thus, inclusion of one extra generation improves the
quality of the fit (compare $\chi^2/n_{d.o.f.} = 23.8/13$ for the
SM from \cite{3b} and $\chi_{\rm min}^2/n_{d.o.f.} = 20.6/12$ from Fig. 3), but it
remains pretty poor.
 If one
multiplies experimental errors of $A_{FB}^b$ and $A_{FB}^c$ by a
factor 10, one gets good quality of SM fit \cite{Ch, 3b} but with
extremely light
higgs, having only a small (few percent) likelihood to be consistent with
the lower limit from direct searches.
The fourth generation allows to have
higgs as heavy as 500 GeV with a perfect quality of the fit:
$\chi_{\rm min}^2/n_{d.o.f.} = 13/12$, if one uses old NuTeV data
(see caption of Fig. 4). Captions of Figs. 3 and 4 reflect also the recent change in NuTeV data
(from $m_W = 80.26 \pm 0.11$ GeV
\cite{10b} to $m_W = 80.14 \pm 0.08$ GeV \cite{11b}) which results
in drastic worsening of the fit even in the presence of the
fourth generation.

To qualitatively understand the dependence of $m_U -m_D$ on
$m_H$ in the case of $m_E = 100$ GeV at $\chi_{\rm min}^2$ let us
recall how radiative
corrections to the ratio $m_W/m_Z$ and to $g_A$ and $R=g_V/g_A$ (the axial and the
ratio of vector and axial couplings of $Z$-boson to charged
leptons) depend on these quantities \cite{9b}:
\begin{equation}
\delta V^i\approx\left[-\left(
\begin{array}{c}
\frac{11}{9}s^2 \\ s^2 \\ s^2 +\frac{1}{9}
\end{array}
\right)\ln(\frac{m_H}{m_Z})^2 +\frac{4}{3}\frac{(m_U
-m_D)^2}{m_Z^2}+\left(
\begin{array}{c}
\frac{16}{9}s^2\frac{m_U -m_D}{m_U + m_D} \\ 0 \\ \frac{2}{9}
\frac{m_U -m_D}{m_U + m_D}
\end{array}
\right)\right] \label{8}
\end{equation}
where $i=m, A, R$, while $s^2 \simeq 0.23$. Corrections to other
observables can be calculated in terms of $\delta V^i$. In the
vicinity of $\chi_{\rm min}^2$ the third term in brackets is much
smaller than the second one. Hence the smallness of the left-right
asymmetry of the plots of Figs. 1, 2.  Since $\frac{11}{9}s^2
\approx s^2 +\frac{1}{9} \approx s^2$, the increase of $m_H$ is
compensated by increase of $|m_U -m_D|$ and we have a valley of
$\chi_{\rm min}^2$.

In conclusion I'd like to make two remarks.

1) Note that the often used parameters $S, T,
U$ (introduced in \cite{Pes})are not adequate for the above analysis, because they assume
that all particles of the fourth generation are much heavier than
$m_Z$, while in our case the best fit corresponds to $m_N \sim
m_Z/2$. In the paper \cite{4} modified definitions of $S$ and $U$
were used in order to deal with new particles with masses
comparable to $m_Z$. However, both original and
modified definitions of $S$, $T$ and $U$ take into account
radiative corrections from the ``light'' 4th neutrino only
approximately, while the threshold effects, that are so important
for $m_N \simeq 50$ GeV, can be adequately described in the
framework of functions $V^i$ as it was done in ref.  \cite{3a, 3b}. 
( For narrow region $m_Z/2 < m_N < 46.5$ GeV (that is actually excluded by direct LEPII data 
on the reactions $e^+ e^- \to \gamma + \nu \bar\nu, \gamma +N \bar N$) 
the threshold effects are so large that modify the Breit-Wigner shape of $Z$-line.
To describe this region one needs different formalism.)

2) Note that in the framework of SUSY with
three generations radiative corrections due to loops with
superpartners also shift upward the mass of the higgs in the case
of not too heavy squarks (300-400 GeV, see Table 1
 in  \cite{gaidaenko1999}) or light
sneutrinos (55-80 GeV, see  \cite{alt2001}).

I am grateful to the organizers of La Thuile conference,
particular to Mario Greco, for their outstanding hospitality and
for excellent conference. I am also grateful to my coauthors
 L.Okun, A.Rozanov, and M.Vysotsky for the pleasure to discuss this piece of Physics.
This work was partly supported by the grant INTAS OPEN 2000-110.

\end{document}